\documentclass[conference,letterpaper,10pt]{IEEEtran}

\usepackage{graphicx}
\usepackage{pgfplots}
\pgfplotsset{compat=1.6}
\usepackage{subcaption}
\usepackage{pgfplotstable}
\usepackage{tikz}
\usepackage{algorithm}
\usepackage{algpseudocode}
\usepackage{mathtools}
\usepackage{multirow}

\begin{document}

\title{Energy-Efficient Mechanism for Smart Communication in Cellular Networks}

\newcommand{\orcidauthorA}{0000-0000-000-000X} 

\author{
\IEEEauthorblockN{ Syed Waqas Haider Shah\IEEEauthorrefmark{1}, Ahmad Talal Riaz\IEEEauthorrefmark{2}, Zahoor Fatima\IEEEauthorrefmark{3}}
\IEEEauthorblockA{\IEEEauthorrefmark{1}Department of Electrical Engineering, Information Technology University, Lahore, Pakistan\\ \IEEEauthorrefmark{2}Experts Vision Engineering and Technology Innovations, Islamabad, Pakistan \\
\IEEEauthorrefmark{2}Department of Electrical Engineering, Army Public College of Management and Sciences, Rawalpindi, Pakistan \\\ waqas.haider@itu.edu.pk, talal@eveati.com, fatima.151214@gmail.com}
}

\maketitle

\begin{abstract}
Internet-of-things (IoT) is gaining popularity in recent times. Cellular network can be an appropriate solution for the IoT connectivity. With the exponential increase in the number of connected IoT devices, the energy consumption is a bottleneck for wireless connectivity solutions. In this work, we proposed a scheme to decrease the power consumption of a base station of a cell. The maximum power consumption of a base station is associated with its data processing, routing, and transmission. Our scheme reduce the size of the transmission data which ultimately reduce the power consumption associated with it. Multiple data reduction techniques have been used for this purpose. System level simulations shows the efficiency of our proposed system.
\end{abstract}



\section{Introduction}
According to a latest UN report, 68\% of the world's population would be living in urban areas by 2050\cite{Esa.un.org. (2018)}. This outlook accentuates the need for smarter cities: cities which would use the latest information and communication technologies (ICT) to improve the economic and social aspects of  mammoth populations within minimum operational cost and optimum utilization of resources. One solution to the burgeoning smart cities is presented by the notion of Internet-of-Things (IoT), which transforms the idea of a city into a network of interconnected variety of nodes, encompassing sensors, actuators, vehicles and controllers along with the computational and analysis units to process the huge volumes of data\cite{Zanella2014}. A smart city consists of various sub-systems such as smart grids, smart home, intelligent energy management systems and smart traffic management\cite{Bellavista2013}.

Communication plays a major role in the Internet of Things, and also defining mobile communication as one of the main platforms for IoT \cite{atzori2010} since the number of connected devices will be up to 100 billion by 2030 \cite{strategy2015}. This would drastically increase the contribution of mobile communication in net global energy consumption and therefore increasing not only the operational expenditures (OPEX) but also contributing to the global rise in carbon footprint. Hence, it's imperative that more energy-efficient and environment friendly networks be developed and deployed to mitigate the environmental threats in smart cities.
This has opened the gates for a new area of research in the recent past, for which the coined term is `greener cellular networks'. As evident, apart from converging to perfection in data rates spectrum efficiency and delays, the new researches have also encompassed environmental issues as well as minimizing the energy consumption\cite{gandotra2017}. Recently, more research studies have encompassed the energy aspect of the mobile communication networks due to its substantive share in the carbon footprint, equivalent to almost 8 million cars per mobile network\cite{Krishnamachari2011}.

\begin{figure}[H]
\centering
\includegraphics[width=3.2in]{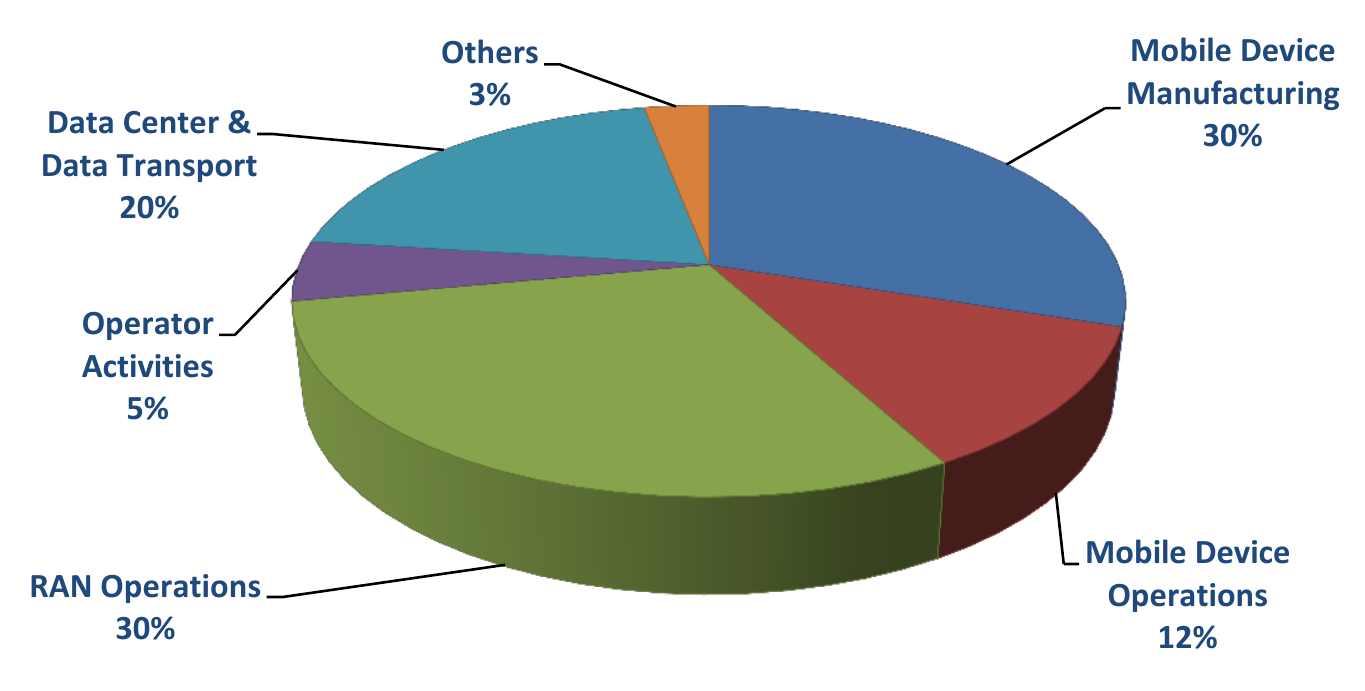}
\caption{Carbon footprint of mobile communication.}
\end{figure}

Typically, the biggest chunk of energy is consumed by the base stations (BS) in mobile networks, accounting for around 60-80\% of the network's total energy consumption\cite{Marsan2009}. Furthermore, a base station devours 90\% of its peak power even in the case of no traffic\cite{Krishnamachari2011}.
Nonetheless, more base stations are being deployed by the network operators in order to cope with the ever-increasing traffic demand. Therefore, an energy efficient solution is required to minimize the energy consumption and carbon footprint along with the reduction in the OPEX of the mobile operators.
\subsection{Contribution and Organization}
In this paper, we have presented an efficient solution for reducing the energy consumption by introducing a base station monitoring unit (BSMU) consisting of a `Hard Unit', which samples, computes and transmits data using sensors and micro-controllers along with an intelligent unit which reduces the size of the data to be transmitted to the distributed cloud. The intelligent unit incorporates HDFS and MapReduce algorithms to significantly reduce the file size and therefore less energy is expended in the Base Station. The Hadoop Distributed File System (HDFS) makes it possible to process the large dataset by breaking it into smaller parts and then using commodity hardware such as commonly available low-cost servers\cite{Borthakur2008}. MapReduce\cite{Dean2008} is an efficient tool for processing large datasets as well as producing large datasets, since many real world tasks can be modeled. The data is compressed at the MapReduce and sent to the distributed cloud for storage. The cloud can coordinate with the individual user devices via the BS to send important notifications.
\begin{table*}[h]
\caption{Techniques for Energy Efficiency at Network Level}
\label{table1}
\centering
\begin{tabular}{|c|c|c|c|}
\hline
\textbf{Techniques}           & \textbf{Advantages}                                                                                     & \textbf{Disadvantages}                                                                                                      & \textbf{Literature}                                                                  \\ \hline
BS switch-off                 & \begin{tabular}[c]{@{}c@{}}High energy-efficient\\ Easy to implement on hardware level\end{tabular}     & \begin{tabular}[c]{@{}c@{}}Effects QoS\\ Need proper optimization\end{tabular}                                              & \cite{Krishnamachari2013,Zhou2013,Ashraf2011,Frenger2011,Chang2017} \\ \hline
Network densification         & \begin{tabular}[c]{@{}c@{}}Energy-efficient \\ Increase system capacity\end{tabular}                    & \begin{tabular}[c]{@{}c@{}}Location of nodes must be\\ calculated acutely probably\\ using stochastic geometry\end{tabular} & \cite{Kountouris2014, Leem2010,Kelif2011}                           \\ \hline
Massive MIMO                  & Increase diversity                                                                                      & \begin{tabular}[c]{@{}c@{}}Increase overall complexity\\ of the system\end{tabular}                                         & \cite{Larsson2014,Larsson2013,Hoydis2013}                           \\ \hline
Efficient resource allocation & \begin{tabular}[c]{@{}c@{}}Optimal utilization of resources\\ Easy to implement\end{tabular}            & \begin{tabular}[c]{@{}c@{}}Relies on trade-offs with\\ other metrics\end{tabular}                                           & \cite{Zhang2012, Zou2013,Singh2017,Huang2015}                       \\ \hline
Network offloading            & \begin{tabular}[c]{@{}c@{}}High energy-efficient\\ Reduces delay\\ Releases burden from BS\end{tabular} & \begin{tabular}[c]{@{}c@{}}Security and interference\\ issues\end{tabular}                                                  & \cite{Boccardi2014,Tehrani2014,Wang2014}                            \\ \hline
\end{tabular}
\end{table*}
\section{Related Work}

\subsection{Power Optimization at Network Level}
In the past, while the main focus of mobile communication was on the enhancement of spectrum efficiency, network capacity and coverage, the recent trend has also shifted to the energy consumption and reduction of carbon footprint due to global warming and exhaustion of natural resources\cite{Bianzino2012}. The academic and industrial research circles have concurred to the rule of increasing the system capacity by 1000x while maintaining at least the current consumption of energy\cite{Andrews2014}.

In terms of mobile communications, there are various fronts at which energy conservation can be applied to reduce the overall carbon front. For example, most of the energy in mobile network is consumed by the BS, in the form of power amplifiers, air-conditioners, RF circuitry etc.  The measure of energy consumed by each of the components in the (BS is depicted in Fig. 2. \cite{gandotra2017}. Therefore, controlling the energy consumption at the BS is an intuitive approach, as targeted by many studies in different ways. For example,  \cite{Krishnamachari2013} have introduced a switching-on/off based energy saving (SWES) scheme to switch off the redundant BS in the network based on current network traffic, while monitoring the additional load on neighboring BS's that might accompany this dynamic switching. This idea is picked by a variety of different researches such as \cite{Chang2017}, which have named it discontinuous transmission (DTX) and have studied its impact on the spectral and energy efficiency of the overall system. Several other studies have followed similar approaches to monitor the network traffic and conveniently turning off certain BS, which can save energy expended by power amplifiers and air-conditioning units including \cite{Zhou2013,Ashraf2011,Frenger2011}.

The switch off mechanism of BS's is easy to implement in the current architecture of mobile networks in which the trend is being shifted towards smaller BS. However, switching off the BS might also result in impact Quality of Service (QoS) because of the decreased system capacity and burdened neighbors unless proper optimization is done and therefore sensitive to non-regularities\cite{Soh2013,Tabassum2014}. Furthermore, by employing denser networks with optimized smaller cell size also contributes to reduction in energy consumption along with increase in system capacity as corroborated by several studies including \cite{Leem2010,Kelif2011}. However, while increasing the density of the BS's saves energy, one important aspect that must be taken into consideration is their location, which has been done using the stochastic geometry as studied in \cite{Weber2012,Kountouris2014}.

\begin{figure}[H]
\centering
\includegraphics[width=3.2in]{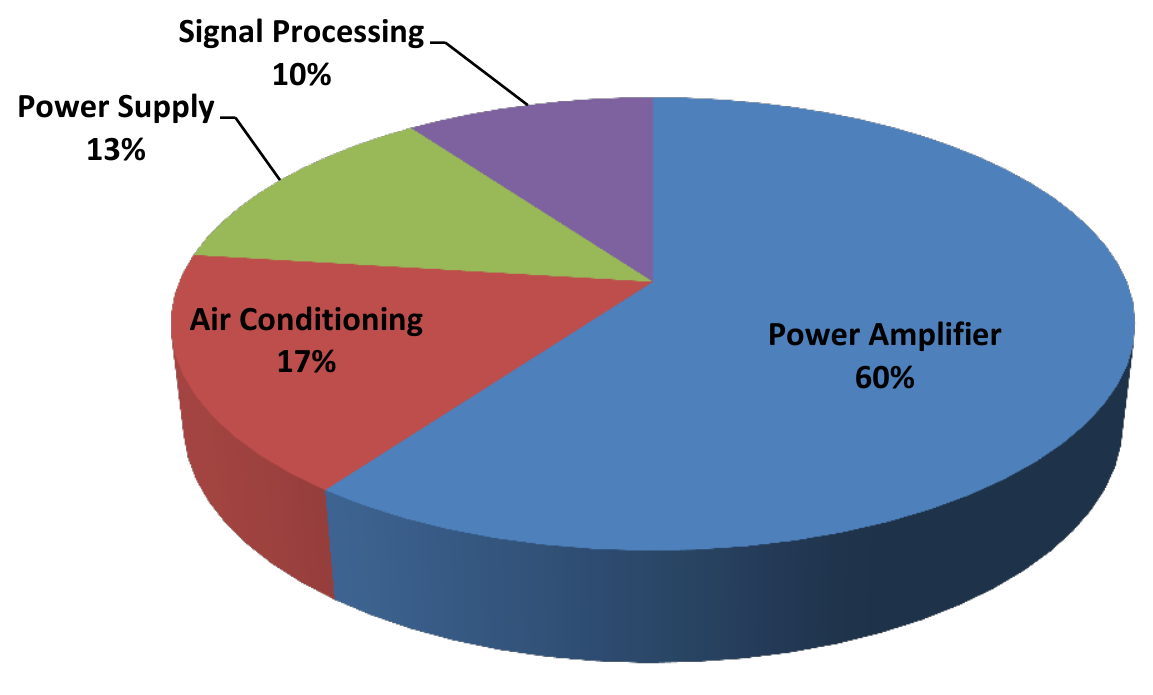}
\caption{Energy consumption by BS components.}
\end{figure}

Another related way to boost up energy efficiency while reducing inter-user interference in mobile networks is to use massive MIMO, in which conventional high-power antennas are substituted by a vast array of smaller and low power antennas as observed by \cite{Larsson2014}. It has been shown in \cite{Larsson2013} that massive MIMO can reduce the radiated power through the multiple antennas by the factor of square root of total number of antenna for the given data rate. However, massive MIMO entails its own disadvantages, first of all being the required complexity of their deployment, followed by difficulties in channel estimation for each antenna as discussed in \cite{Hoydis2013}.

Like massive MIMO, several other approaches can be used to enhance energy efficiency by better utilization of the resources without needing to overhaul the whole system hardware. For example, \cite{Zou2013} tenders the idea of `inter-network' cooperation between the user devices to help transmit the data packets to the BS. However, this can save energy when the two cooperating nodes are closer to each other as compared to the BS. Efficient resource allocation is another way of saving energy which has remained a hot topic in the past, such as proposed by \cite{Singh2017}.Similarly, \cite{Huang2015} has presented the idea of opportunistically availing the unused spectrum, while to save energy the authors derive a maximum independent set (MIS) optimization problem based on transmission power, joint frequency allocation etc. However, all these techniques rely on trade-offs between energy efficiency and other metrics such as bandwidth, delay etc\cite{Zhang2012}.

Yet another energy efficient mechanism is the called the `offloading': shifting the cellular traffic to other radio access technologies (such as WiFi, Bluetooth), whenever possible and convenient. One important application of offloading is device-to-device (D2D) communication, which allows co-located devices to communicate directly with each other, thus reducing the energy consumption drastically, since the BS would no more be the intermediary between the two devices and hence creating room for other devices. Several studies have focused on this aspect of offloading such as \cite{Boccardi2014,Tehrani2014,Wang2014}. One of the major challenges in D2D communication is the proper interference management between the devices, since in certain conditions, the BS does not provide mediation in call setup and resource allocation, which can lead to severe interference and hence a smart interference management is inevitable. Another important challenge is the security concerns when two devices via other intermediary devices, which has serious ramification unless proper encryption protocol is agreed between the two devices\cite{Tehrani2014}. Table \ref{table1} provides a comparison of different techniques for energy-efficiency at network level.

\section{System Model}
The system model is shown in Fig. 3. The system contains three parts, data acquisition using IoT devices and sensors, data analytic and smectic analysis section, and a cloud for further data processing.

\begin{table*}[h]
\caption{Values for the Data Analytic and Semantic Analysis Section}
\label{table2}
\centering
\begin{tabular}{|c|c|c|c|c|}
\hline
\textbf{}                                            & \textbf{Counter}                                                             & \textbf{Original}  & \textbf{Reduction} & \textbf{Total} \\ \hline
\multirow{3}{*}{\textbf{Proposed Mechanism}}      
                                                     & VM Size (KBs)   & 3,755,327 & 1,890,803   & 5,646,131  \\ \cline{2-5}
                                                     & PM Size (KBs)   & 398,012   & 107,339     & 505,352   \\ \cline{2-5}
                                                     
                                                     & Time Elapsed (ms)                                                          & 470           & 990             & 1,460          \\ \cline{2-5}
                                                     & Total (KBs) & 336,592   & 92,798      & 429,391    \\ \hline
\multirow{3}{*}{\textbf{Input Files}} & read (KBs)                                                                  & 5         & 0               & 5          \\ \cline{2-5}
                                                     & read (KBs)                                                              & 5         & 0               & 5          \\ \cline{2-5}
                                                     & Write (KBs)                                                         & 122       & 62          & 99        \\ \hline
\multirow{2}{*}{\textbf{System Files}}       & read (KBs)                                                            & 0             & 0.5             & 0.5            \\ \cline{2-5}
                                                     & Write (KBs)                                                         & 0             & 0.3             &0.3            \\ \hline
\textbf{Output Files}                          & Write (KBs)                                                                & 0             & 0.3             & 0.3            \\ \hline
\end{tabular}
\end{table*}

\begin{figure}[H]
\centering
\includegraphics[width=3.2in]{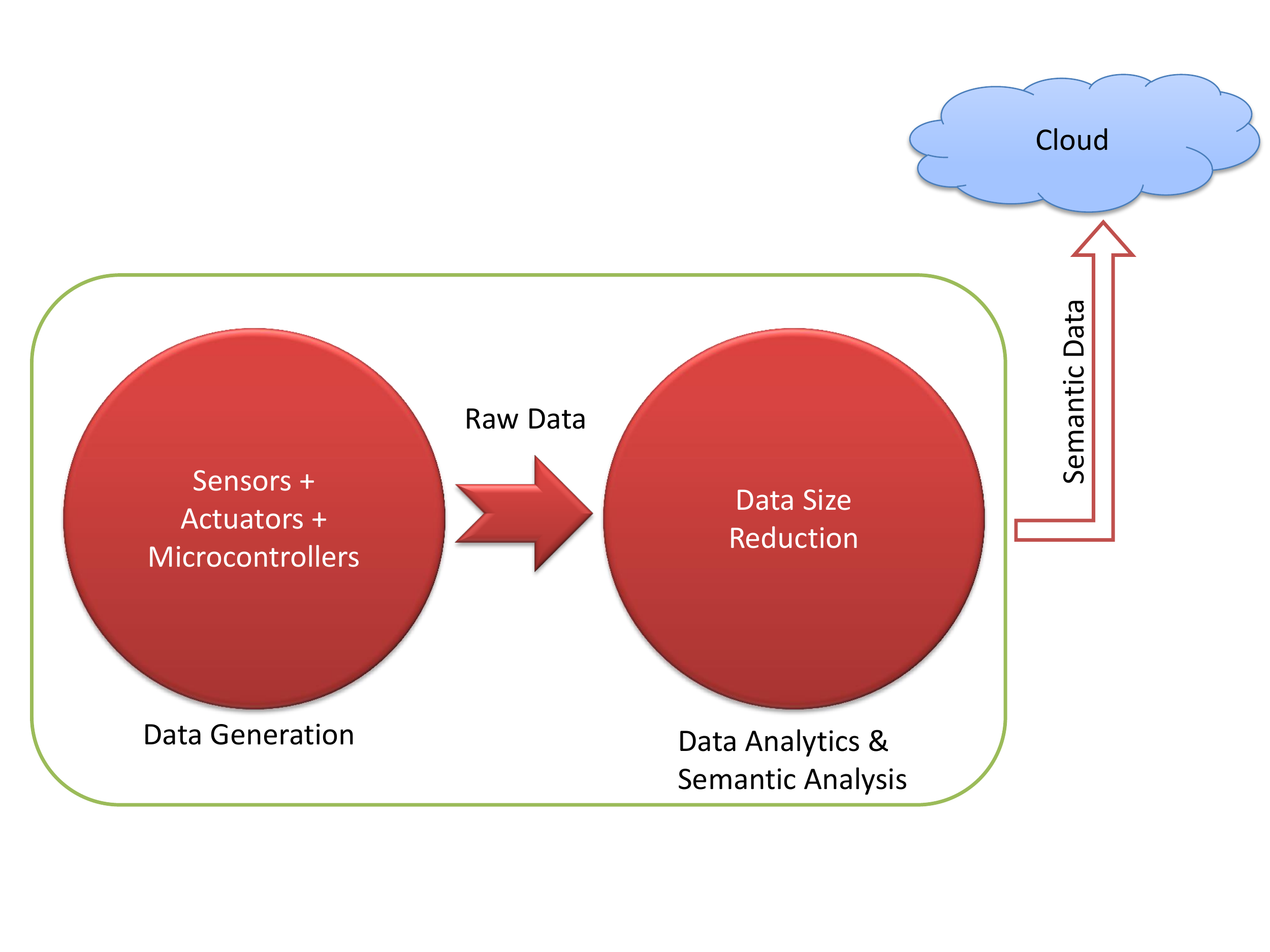}
\caption{In depth view of the proposed system model.}
\label{figureee}
\end{figure}

\section{Simulation Section}
In this section, we perform the simulations according to our system model presented in the Fig. \ref{figureee}. Moreover, the simulation results are presented in the Table \ref{table2} and Table \ref{table3}.

\begin{table*}[h]
\caption{Power Consumption of the Proposed Mechanism}
\label{table3}
\centering
\begin{tabular}{|c|c|c|c|c|}
\hline
\multirow{2}{*}{\textbf{IoT Sensors}} & \multicolumn{2}{c|}{\textbf{File Size}}                                                                                                                           & \multicolumn{2}{c|}{\textbf{Energy Consumption}}                                                                                            \\ \cline{2-5}
                                 & \textbf{\begin{tabular}[c]{@{}c@{}}Before Data Analytics \\ (bytes)\end{tabular}} & \textbf{\begin{tabular}[c]{@{}c@{}}After Data Analytics \\ (bytes)\end{tabular}} & \textbf{\begin{tabular}[c]{@{}c@{}}Traditional Way\\  (mW)\end{tabular}} & \textbf{\begin{tabular}[c]{@{}c@{}}With Proposed Solution \\ (W)\end{tabular}} \\ \hline
\textbf{Moisture}                & 5000                                                                                & 500                                                                                 & $7.99x10^{-3}$                                       & $0.21x10^{-3}$                                \\ \hline
\textbf{Diesel Level}              & 5234                                                                                & 745                                                                                &$ 8.654x10^{-3}$                                       & $2.56x10^{-3}$                                    \\ \hline
\textbf{Smoke}                   & 3975                                                                                & 478                                                                                & $6.78x10^{-3} $                                      & $1.24x10^{-3} $                                  \\ \hline
\textbf{Temperature}              & 5421                                                                                & 324                                                                                & $9.21x10^{-3}$                                       & $1.23x10^{-3}$                                    \\ \hline
\textbf{Pressure}              & 4951                                                                                & 415                                                                                & $8.96x10^{-3}$                                       & $1.87x10^{-3}$                                    \\ \hline
\end{tabular}
\end{table*}

\section{Conclusion}
Internet-of-things is gaining popularity in recent times in almost every sector of the society. Cellular network can be an appropriate solution for the IoT connectivity. With the exponential increase in the number of connected IoT devices, the energy consumption is a bottleneck for wireless connectivity solutions. In this work, we proposed a scheme to decrease the power consumption of a base station of a cell. The maximum power consumption of a base station is associated with its data processing, routing, and transmission. Our scheme reduce the size of the transmission data which ultimately reduce the power consumption associated with it. Multiple data reduction techniques have been used for this purpose. System level simulations shows the efficiency of our proposed system.

\appendices
\footnotesize{
\bibliographystyle{IEEEtran}

}

\end{document}